\documentclass[10pt]{article}
\usepackage{graphicx,cite,times,bibspacing}

\newcommand{\twid}{\sim}

\newcommand{\ddt}{ {d \over dt} }

\newcommand{\av}[1]{\left<\ #1 \ \right>}

\newcommand{\paren}[1]{\left( #1 \right)}
\newcommand{\square}[1]{\left[ #1 \right]}
\newcommand{\curly}[1]{ \left\{ #1 \right\} }

\newcommand{\casesbracketsii}[4]
{\left\{
\begin{array}{ll}
#1\   & (#2) \\ & \blank #3\  & (#4) 
\end{array}%
\right.
}

\newcommand{\casesbracketsshortii}[4]
{\left\{
\begin{array}{ll}
#1\ & (#2) \\  #3\ & (#4) 
\end{array}%
\right.
}

\newcommand{\gap}{\hspace{.4in}}

\newcommand{\ggap}{\hspace{.8in}}

\newcommand{\blank}{\ \\}

\newcommand{\gt}{\rightarrow}

\newcommand{\period}{\ \ .}
\newcommand{\comma}{\ ,\ }

\newcommand{\ignore}[1]{}

{\end{list}}

{\end{list}}

{\ \\ XXXXX \hfill XXXXX #1 XXXXX\begin{equation}\label{#1}}%
{\end{equation}}

{\ \\ XXXXX \hfill XXXXX #1 XXXXX\begin{eqnarray}\label{#1}}%
{\end{eqnarray}}

\newenvironment{bequation}[1]%
{\begin{equation}\label{#1}}%
{\end{equation}}

\newenvironment{beqnarray}[1]%
{\begin{eqnarray}\label{#1}}%
{\end{eqnarray}}

\newcommand{\drop}{\nonumber \\}

\newcommand{\ddrop}{\drop\drop}

\newcommand{\ie}{i.\,e.~}
\newcommand{\eg}{e.\,g.\@ }

{\begin{center} \section*{#1} \end{center}}

{\end{figure}} 

{\end{figure}}

{\end{figure}}

\newenvironment{eq}[1]%
{\begin{bequation}{#1}}{\end{bequation}}

\newenvironment{eqarray}[1]%
{\begin{beqnarray}{#1}}{\end{beqnarray}}

\def\eqref#1{(\ref{#1})}

\newcommand{\ccrit}{c_{\rm crit}}

\newcommand{\kpluseff}{k^+_{\rm eff}}

\newcommand{\rH}{r_{\rm H}}
\newcommand{\rPi}{r_{\rm Pi}}

\newcommand{\kplusT}{k^{+}_{\rm T}}

\newcommand{\vminusP}{v^{-}_{\rm P}}

\newcommand{\vminusD}{v^{-}_{\rm D}}

\newcommand{\vminusT}{v^{-}_{\rm T}}
\newcommand{\vminusTD}{v^{-}_{\rm T|D}}

\newcommand{\tcap}{t_{\rm cap}}
\newcommand{\Ncap}{N_{\rm cap}}
\newcommand{\Ncapcrit}{N_{\rm cap}^{\rm crit}}
\newcommand{\NATPcap}{N_{\rm cap}^{\rm ATP}}
\newcommand{\NADPPicap}{N_{\rm cap}^{\rm ADPPi}}
\newcommand{\Dcap}{D_{\rm cap}}
\newcommand{\vcap}{v_{\rm cap}}

\newcommand{\Dinf}{D_\infty}

\newcommand{\Poisson}{{\cal P}}

\newcommand{\psiT}{\psi^{\rm T}}
\newcommand{\psiD}{\psi^{\rm D}}
\newcommand{\psiP}{\psi^{\rm P}}

\newcommand{\phiT}{\phi_{\rm T}}
\newcommand{\phiD}{\phi_{\rm D}}
\newcommand{\phiP}{\phi_{\rm P}}

\newcommand{\PsiT}{\Psi^{\rm T}}
\newcommand{\PsiD}{\Psi^{\rm D}}
\newcommand{\PsiP}{\Psi^{\rm P}}

\newcommand{\pcore}{p_{\rm core}}
\newcommand{\Sinf}{S_\infty}

\begin{document}

\pagenumbering{arabic}

\twocolumn



\renewcommand{\today}{}

\title{\bf \LARGE Actin Polymerization Kinetics, Cap Structure and
Fluctuations} 

\author{ {\bf Dimitrios Vavylonis$^\dagger$, Qingbo Yang$^\ddagger$,
and Ben O'Shaughnessy$^\dagger$ \footnote{To whom
correspondence should be addressed. E-mail: bo8@columbia.edu}
}\\
{\small Departments of $^\dagger$Chemical Engineering and
$^\ddagger$Physics, Columbia University, New York, NY 10027, USA 
} 
}

\maketitle

{\bf Polymerization of actin proteins into dynamic structures is
essential to eukaryotic cell life.  This has motivated a large body of
in vitro experiments measuring polymerization kinetics of individual
filaments.  Here we model these kinetics, accounting for all relevant
steps revealed by experiment: polymerization, depolymerization, random
ATP hydrolysis and release of phosphate (Pi).  We relate filament
growth rates to the dynamics of ATP-actin and ADP-Pi-actin caps which
develop at filament ends.  At the critical concentration of the barbed
end, $\ccrit$, we find a short ATP cap and a long
fluctuation-stabilized ADP-Pi cap.  We show that growth rates and the
critical concentration at the barbed end are intimately related to cap
structure and dynamics.  Fluctuations in filament lengths are
described by the length diffusion coefficient, $D$.  Recently Fujiwara
et al.  [{\em Nature Cell Biol.}  (2002) {\bf 4}, 666] and Kuhn and
Pollard [{\em Biohys.  J.}  (2005) {\bf 88}, 1387] observed large
length fluctuations slightly above $\ccrit$, provoking speculation
that growth may proceed by oligomeric rather than monomeric on-off
events.  For the single monomer growth process we find that $D$
exhibits a pronounced peak below $\ccrit$, due to filaments
alternating between capped and uncapped states, a mild version of the
dynamic instability of microtubules.  Fluctuations just above $\ccrit$
are enhanced but much smaller than those reported experimentally.
Future measurements of $D$ as a function of concentration can help
identify the origin of the observed fluctuations.  }


The tendency of actin protein to spontaneously polymerize into rapidly
growing filaments is fundamental to the life of eukaryotic cells.
Cell motility\cite{bray:motility_book,pollardborisy:motility_review_2003}, cell division\cite{pelhamchang:actin_dynamics_cytokinesis}, 
and endocytosis\cite{young:yeast_patch_dendritic_jcb_04} are examples of
processes exploiting the dynamic character of actin structures
composed of filaments. The regulation of
filament growth processes leads to well-defined structures and
coordinated function.  For example, in combination with branching,
capping, and depolymerizing proteins, actin self-assembles into
controlled dynamic cross-linked networks forming the dynamic core of
lamellipodia\cite{pollardborisy:motility_review_2003}.

\begin{figure}[tb]
\centering
\includegraphics[width=8cm]{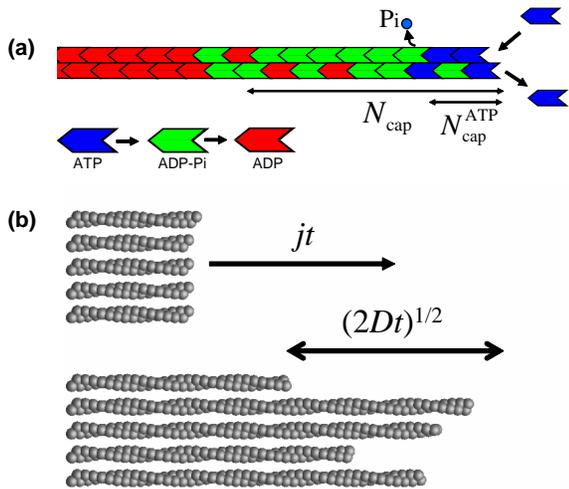}
\caption{\label{actin_scheme} \footnotesize 
(a) Schematic of the 3-species cap at the barbed end of a long actin
filament.  Near the critical concentration we find a
fluctuation-induced cap of $\Ncap \approx 25$ monomers, with a short
ATP-actin component, $\NATPcap$ of order one.  (b) Mean growth rate
and fluctuations: in time $t$ the average number of monomers added to
a filament end is $jt$, with a spread of $(2Dt)^{1/2}$ about this value.}
\end{figure}

These complex cellular actin-based systems exhibit multiple superposed
mechanisms. This has inspired a large body of in vitro work aiming to
unravel these mechanisms and pin down rate constants for the
constituent processes in purified systems
\cite{korn:actin_review_science}. An important class of
experiments entails measuring growth rate at one end by microscopy
\cite{pollard:actin_rate_constants,amannpollard:arp23_branch_real_time_viz,fujiwara:actin_microscopy_large_fluc,kuhnpollard:tirf_bj_04}
or by bulk spectroscopic methods
\cite{carlier:atp_cap_evidence,carlier:adp_atp_sonication,
couekorn:hydro_pointed_end,carlier:cytochalasin, carlier:effect_of_Mg,
weber:actin_kinetics_pointed_end,carlier:adppi_role} as a function of
actin monomer concentration.  From these and other in vitro studies
using various labeling techniques the following picture has emerged of
filament growth kinetics in the presence of ATP (see fig.
\ref{actin_scheme}). (i) Monomers are added to a growing filament end
as ATP-actin.  (ii) Rapidly, the ATP is then hydrolyzed to ADP and
phosphate (Pi), both remaining bound to the monomer host
(ADP-Pi-actin)\cite{pardeespudich:actin_assembly_mechanism_82,pollardweeds:atp_hydro_rate,
carlier:atp_cap_evidence,carlier:hydrolysis_mechanism,
carlier:effect_of_Mg,ohmwegner:actin_hydro_cap,
pieperwegner:atp_cap_struct,
blanchoinpollard:atp_hydro_quenched_flow}.  A rate of $0.3$s$^{-1}$
was reported in ref.
\citenum{blanchoinpollard:atp_hydro_quenched_flow} in the presence of
Mg, assuming random hydrolysis uninfluenced by neighboring monomers.
(iii) After a long delay Pi release into solution occurs,
generating ADP-actin
\cite{carlierpantaloni:pi_release_direct_evidence,
carlier:pi_release_rate_linked_enzyme,
melki:carlier:continuous_monitor_pi}.  Reported release rates are in
the range $0.002-0.006$s$^{-1}$
\cite{carlierpantaloni:pi_release_direct_evidence,
carlier:pi_release_rate_linked_enzyme,
melki:carlier:continuous_monitor_pi,blanchoin:pollard_adf_arp23_cap_currbio_00}.

A typical filament in a growth rate experiment is thousands of monomer
units in length and thus consists mainly of ADP-actin.  Hence the
picture which emerges is of a long ADP-actin filament with a complex
3-state {\em cap} region at the filament end
\cite{korn:actin_review_science} (see fig. \ref{actin_scheme}).  A
major goal of this report is to establish the composition and kinetics
of the cap, and how these determine growth rates and measurable length
fluctuations. This is important in the context of cellular processes:
the monomer composition in actin filaments is thought to regulate
actin-binding proteins in a timely and spatially organized way
\cite{pollardborisy:motility_review_2003}.  For example, it has been
suggested that rates of branching generated by the Arp2/3 protein
complex and/or debranching processes may depend on which of the 3
monomer species is involved, ATP-actin, ADP-Pi-actin or ADP-actin
\cite{blanchoin:pollard_adf_arp23_cap_currbio_00,
ichetovkin:cofilin_new_dendritic_02,amannpollard:arp23_branch_real_time_viz}.
Pi release has been proposed to act as a timer for the action of the
depolymerizing/severing protein ADF/cofilin which preferentially
attacks ADP-actin \cite{pollardborisy:motility_review_2003}.

Our aim in this report is to establish theoretically the quantitative
implications of the currently held picture of actin polymerization.
\ignore{
 We will argue that some features of filament kinetics are
universal, whereas others depend on detailed numerical parameter
values.  } 
Previous
theoretical works addressed growth rates before the important process
of Pi release was established
\cite{pantaloni:hill_actin_atp_cap_1,hill:actin_atp_cap_2,hill:aggr_bio_book}.
To date there has been no theoretical analysis of single filament non
steady state growth rates rigorously accounting for the processes
(i)-(iii) above.  A recent theoretical work
\cite{bindschadler:actin_mechanistic_model} has addressed steady state
filament compositions.


The cap has important consequences for the growth rate $j$ as a
function of ATP-actin concentration, $c$.  Measured $j(c)$ curves,
such as those in fig.  \ref{jc_exp}, are strikingly non-linear in the
region near the concentration where growth rate vanishes
\cite{carlier:adppi_role,keiser:actin_growth_nonlinear}.  These become
almost linear in excess Pi studies, where presumably the ADP-actin
species is no longer involved \cite{carlier:adppi_role}.  The
complexity of the cap structure and dynamics also underlies the values
of the critical concentration $\ccrit$ at the fast-growing ``barbed''
end and slow-growing ``pointed'' end of the polar actin filament
($\ccrit$ denotes the concentration where mean growth rate at one end
vanishes.)  It is well known that in general these are different since
detailed balance cannot be invoked for these non-equilibrium polymers
\cite{hill:aggr_bio_book}.  Our work explores how these differences
are related to cap structure.

The major experimental focus has been mean growth rates, $j(c)$.
However, equally revealing are {\em fluctuations} about the mean whose
measurement can expose features of the dynamical processes occurring
at filament ends unavailable from $j(c)$.  These fluctuations are
characterized by a ``length diffusivity'' $D$ measuring the spread in
filament lengths (see fig. \ref{actin_scheme}(b)) similarly to simple
one dimensional Fickian diffusion: after time $t$, the root mean
square fluctuation in filament length is $(2 D t)^{1/2}$ about the
mean value $j(c) t$.  Using single filament microscopy, Fujiwara et
al.  \cite{fujiwara:actin_microscopy_large_fluc} and Kuhn and Pollard
\cite{kuhnpollard:tirf_bj_04} recently measured unexpectedly high
values of this diffusivity near steady state conditions, $D \twid
30$mon$^2$/s.  This should be compared with what would be expected of
an equilibrium polymerization involving the measured on/off rates of
order 1mon/s, which would lead to $D \twid 1$mon$^2$/s
\cite{fujiwara:actin_microscopy_large_fluc,ben:living_ionic_letter,ben:living_ionic,
hill:aggr_bio_book}.  A number of possible explanations were proposed.
(i) Fluctuations arise from ``dynamic instability'' due to stochastic
cap loss episodes.  This would be a far milder version of the
``catastrophes'' in microtubule polymerization
\cite{littlefieldfowler:comment_fujiwara_actin_large_fluc,%
fujiwara:actin_microscopy_large_fluc}.  (ii) Filament polymerization
proceeds by addition and subtraction of {\em oligomeric} actin segment
\cite{littlefieldfowler:comment_fujiwara_actin_large_fluc,%
fujiwara:actin_microscopy_large_fluc}; this would constitute a radical
departure from the accepted picture of filament growth kinetics
involving single monomer addition events.  (iii) Growth involves extra
stochastic events such as short pauses possibly originating in
filament-surface attachments \cite{kuhnpollard:tirf_bj_04}.  (iv)
Enhanced fluctuations results from an artifact due to monomer
labeling \cite{kudryashov:rhodamine_destab_bj_04}. (v) Experimental
error in filament length measurements \cite{kuhnpollard:tirf_bj_04}. A
major focus of this report is to calculate the concentration-dependent
length diffusivity, $D(c)$, assuming that the standard
monomer-by-monomer addition picture is valid.  We will see that this
leads to large $D$ values {\em below} $\ccrit$; just above the
critical concentration fluctuations are enhanced, though much less
than the experimental values.

We consider the initial condition where long pre-formed ADP-actin seeds
are exposed initially to a buffer of {\em fixed} actin concentration
$c$ and excess ATP.  Thus for a given $c$ value, a filament consists
of a very long ADP-actin core at the end of which lies a complex
steady state (but fluctuating) ATP-actin/ADP-Pi-actin cap.  Our
analysis emphasizes the barbed end, the pointed end assumed blocked.
Our results apply to very dilute filaments where only ATP-actin is
assumed to add to filaments since (i) free monomers bind ATP more
strongly than ADP \cite{kinosian:actin_nucleotide_binding}, and (ii)
depolymerized ADP-actin or ADP-Pi-actin has enough time to exchange
its nucleotide for ATP before repolymerization.  An important issue is
the nature of the ATP hydrolysis mechanism: the experiments of refs.
\citenum{ohmwegner:actin_hydro_cap,pieperwegner:atp_cap_struct}
support a random mechanism, though others have suggested a cooperative
vectorial mechanism occurring at the interface between ADP-Pi-actin
and ATP-actin with rate $13.6$s$^{-1}$
\cite{carlier:hydrolysis_mechanism,pantaloni:hill_actin_atp_cap_1}.
In this study, random hydrolysis is assumed throughout.


\section*{\large Parameter Values and Mathematical Methods}

One of the major aims of this work is to identify qualitative, but
experimentally measurable, features of the growth kinetics which are
independent of the precise values of rate constants, since the latter
depend on experimental conditions such as ionic strength
\cite{drenckhahnpollard:actin_dc_86} and the values themselves are
often controversial.  The parameter values we use are shown in table 1
in which $\kplusT$ is the polymerization rate constant of ATP-actin,
and $\vminusT$, $\vminusD$ and $\vminusP$ are the depolymerization
rates of ATP-actin, ADP-actin, and ADP-Pi-actin, respectively.  The
rates of ATP-hydrolysis and Pi release (both assumed irreversible) are
$\rH$ and $\rPi$, respectively.  In addition, we will explore the
effects of changing some of these parameter values.  Since the monomer
at the tip makes bonds with two nearest neighbors, each belonging to a
different protofilament, one expects that rate constants may also
depend on the state of neighbors.  Here, however, we study the
simplest ``one body'' model, assuming on/off rates depend only on the
attaching/detaching species\cite{pollard:actin_rate_constants} and
that hydrolysis and Pi release rates are uniform along the filament.
The influence of ``many body'' effects will be briefly discussed
below.

To calculate filament growth kinetics and composition one is faced
with the formidable task of obtaining the steady state probability
distribution of all possible actin monomer sequences along the
filament: there are 3 possible states per monomer, so for filaments of
$N$ units long this necessitates solving $3^N$ coupled equations.  We
have managed, however, to obtain a solution for the mean elongation
rate $j(c)$ by projecting the full system of $3^N$ equations onto a
set of just 3 exact equations for the return probabilities $\psiT_t,
\psiP_t$, and $\psiD_t$.  These are the probabilities that a given
monomer which was polymerized at $t=0$ is again at the tip at time $t$
as ATP-actin, ADP-Pi-actin, or ADP-actin, respectively.

The outline of our method is as follows.  For $j<0$ the growth rate is
related to the return probabilities by $j = \vminusD \pcore$, where
$\pcore = 1-\int_0^\infty dt(\psiD_t + \psiP_t +\psiT_t)$ is the
probability of exposure of the ADP-actin core at the tip.  For $j>0$,
the relation is $j = \kplusT c -\int_0^\infty dt F_t$ where $F_t
\equiv \psiT_t \vminusT + \psiP_t \vminusP + \psiD_t \vminusD$ is the
mean depolymerization rate at time $t$ of a monomer which added to the
tip at $t=0$.  The integral of $F_t$ is the total depolymerization
rate of added monomers.  In the supporting material we present the
dynamical equations obeyed by the return probabilities, from which we
obtained a closed recursion relation for the Laplace transform of
$F_t$, namely $f_E$.  This relates $f_E$ to $f_{E+\rH}$ and
$f_{E+\rPi}$.  With boundary condition $f_E \gt 0$ as $E\gt\infty$, we
started from large $E$ values and evolved this equation numerically
towards $E =0$ to obtain $f_0 \equiv \int_0^\infty dt F_t$.  Given
$f_0$, the time integrals of the return probabilities were then
directly obtained from the dynamical equations and $j$ was thereby
determined.

The above analytically based method does not generate cap sizes and
length diffusivities.  In order to calculate these quantities and also
to test the validity of the analytical method we have simulated the
stochastic tip dynamics employing the kinetic Monte Carlo (MC) method
known as the BKL \cite{bortz:kalos:lebowitz:algorithm_monte_carlo} or
Gillespie \cite{gillespie:exact_stochastic_jphyschem_77} algorithm, to
evolve the state of a filament tip in time and to calculate its mean
growth rate.  Each step of the algorithm entails updating time by an
amount depending on the rate and number of possible future events,
namely polymerization/de\-polymerization, hydrolysis, and Pi release.
Excellent agreement is found between MC results and the numerical
solutions of our closed equations for the growth rate (see inset of
fig. \ref{jc_ADP}).


Our analytical method is exact and avoids preaveraging, an
approximation where the joint probability of a given filament
nucleotide sequence is approximated as a product of probabilities for
individual actin subunits.  This neglects correlations between units.
To assess the accuracy of this scheme, we compared our results for cap
size and growth rate to those obtained using preaveraging (see
supporting material for details).  Preaveraging has been used in
other theoretical studies of actin polymerization such as ref.
\cite{bindschadler:actin_mechanistic_model} to study steady state and
ref.  \cite{keiser:actin_growth_nonlinear} to study growth rates.

\begin{table}
\begin{tabular}{|c|c|c|c|c|c|c|}   \hline
$\kplusT (\mu$M$^{-1}$s$^{-1})$
& $\vminusT$
& $\vminusP$
& $\vminusD$
& $\rH$
& $\rPi$    \\ 
11.6$^{(a)}$
& 1.4$^{(a)}$
& 1.1$^{(b)}$
& 7.2$^{(a)}$
& 0.3$^{(c)}$
& 0.004$^{(d)}$  \\ \hline
\end{tabular}
\caption{\footnotesize 
Values of barbed end rate constants used in this work,
appropriate for solutions of 50mM KCl and
1mM MgCl$_2$.
Units in s$^{-1}$ unless otherwise indicated.
$^{(a)}$ From ref. \cite{pollard:actin_rate_constants}. 
$^{(b)}$ Assigned; at present there is no direct measurement
of $\vminusP$, but $j(c)$
measurements with excess Pi \cite{carlier:adppi_role} show the
sum of the ADP-Pi-actin off rates at both ends together is a few times smaller
than $\vminusD$.
$^{(c)}$ From ref. \cite{blanchoinpollard:atp_hydro_quenched_flow}.
$^{(d)}$  From refs. 
\cite{carlierpantaloni:pi_release_direct_evidence,
carlier:pi_release_rate_linked_enzyme,
melki:carlier:continuous_monitor_pi,blanchoin:pollard_adf_arp23_cap_currbio_00}.
}
\end{table}


\section*{\large Cap Structure and the Importance of Fluctuations}

\begin{figure}[tb]
\centering
\includegraphics[width=8cm]{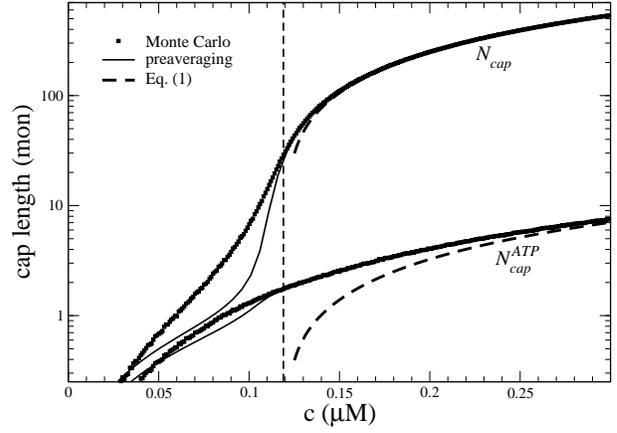}
\caption{\label{cap} \footnotesize 
Total cap length $\Ncap(c)$ and ATP-actin cap length $\NATPcap(c)$ at
barbed end.  Parameters from table 1.  Squares: MC results.  Dashed
lines: eq.  \eqref{bothcaps}. Solid lines: preaveraging approximation.  Vertical dashed line
indicates $\ccrit=0.119 \mu$M.  }
\end{figure}

Using the parameters of table 1, in fig. \ref{cap} we present MC
results for (i) the total cap size, $\Ncap$, namely the mean total
number of ATP-actin and ADP-Pi-actin subunits at the barbed end, as a
function of concentration, and (ii) the number of ATP-actin cap
subunits, $\NATPcap$.  The figure shows that both caps become large
for large concentrations.  This is easy to understand. Consider for
example the ATP cap: when polymerization rates exceed both the
hydrolysis rate $\rH$ and the depolymerization rates, the interface
between ADP-Pi-actin and ATP-actin follows the growing tip with a lag
of $j(c)/\rH$ monomers.  Thus
                                                \begin{eq}{bothcaps}
\NATPcap = j(c)/\rH \comma \  \NADPPicap = j(c)/\rPi \ \ \   (c \gg \ccrit)
                                \period
\end{eq}
Here the number of ADP-Pi subunits, $\NADPPicap$, is found using
similar reasoning as for $\NATPcap$.  The validity of eq.
\eqref{bothcaps} for large concentrations is verified against MC data
in fig.  \ref{cap}.

The striking feature of fig. \ref{cap} is that the total cap remains
long even {\em below} the critical concentration of the barbed end,
being 25 units at $\ccrit$ and remaining larger than unity down to $c
\approx \ccrit/2$.  One might naively have guessed that below $\ccrit$
there would be no cap at all, since the filament is shrinking into its
ADP core.  (Indeed, the absence of a cap would also be suggested by
eq.  \eqref{bothcaps} if one were to extend its validity down to
$\ccrit$ where $j=0$.)  This reasoning is however invalid because it
neglects fluctuations due to randomness of monomer
addition/subtraction.

To understand why fluctuations lead to long caps, consider the length
changes of the cap only, excluding changes in the ADP-actin core
length.  Just below the critical concentration the tip of a typical
long cap has a net shrinkage rate
\cite{ben:living_ionic_letter,ben:living_ionic}, $\vcap(c)$. This is a
weighted average of rates, summed over all possible states of the
short ATP-actin segment on top of the long ADP-Pi-actin segment.
Since $\vcap$ is a smooth function of $c$, it can be Taylor-expanded
near the critical concentration and expressed as $\vcap = \kpluseff (c
- \ccrit)$ where $\kpluseff$ is an effective on rate constant,
different from $\kplusT$.  Now superposed on this average shrinkage
the cap tip also performs a random walk in cap length space, described
by a diffusivity $\Dcap(c)$
\cite{fujiwara:actin_microscopy_large_fluc,ben:living_ionic_letter,ben:living_ionic},
also an average over the states of the short ATP cap.  ($\Dcap$ is in
fact the {\em short-time} diffusivity of the entire filament, see
discussion below.)  For small times, diffusivity dominates and of
order $(2 \Dcap t)^{1/2}$ units add to or subtract from the cap.  For
times less than the cap turnover time $\tcap$, this is much bigger
than the number of units wiped out by coherent shrinkage, $\vcap t$.
The cap lifetime $\tcap$ is the time when the shrinkage just catches
up, $\vcap \tcap \approx (2\Dcap \tcap)^{1/2}$.  Hence the approximate
dependence of cap length on concentration is
                                                \begin{eq}{simple}
\Ncap = \vcap\tcap \approx 2 \Dcap/[\kpluseff(\ccrit - c)] \comma \ \ \ (c < \ccrit)
\end{eq}
which indeed becomes large as $\ccrit$ is approached from below.

In summary, even though {\em on average} below $\ccrit$ no ATP-actin
monomers are being added to the tip, {\em fluctuations} in
addition/subtraction rates allow a cap to grow to length $(2\Dcap
\tcap)^{1/2}$ because the cap length diffusivity is dominant for times
less than $\tcap$.  Now since Pi release is very slow, for simplicity
in deriving eq. \eqref{simple} we assumed the release rate was zero,
$\rPi=0$. However, the result of eq. \eqref{simple} is valid even for
a non-zero $\rPi$ except for concentrations so close to $\ccrit$ that
the cap turnover time exceeds the Pi release time.  In this inner
region, diffusion is only able to grow the cap for a time of order
$\rPi^{-1}$ before Pi release intervenes. The maximum possible cap
length, attained very close to $\ccrit$, is thus
                                                \begin{eq}{cap_crit}
\Ncapcrit \approx [2 \Dcap(\ccrit)/\rPi]^{1/2} \period
\end{eq}
Eq. \eqref{simple} is valid until $\Ncap$ reaches this bound.

These arguments explain the origin of the long caps below $\ccrit$.
To make quantitative comparison of eqs. \eqref{simple} and
\eqref{cap_crit} to the numerics of fig. \ref{cap}, the values of
$\Dcap$ and $\vcap$ must be determined. Now since for our parameter
set $\vminusT$ and $\vminusP$ have similar values (see table 1), an
estimate can be obtained by considering the special case where
$\vminusT=\vminusP$ (identical ATP-actin and ADP-Pi-actin).  This is a
convenient case because $\Dcap$ and $\vcap$ can be calculated exactly;
the cap has just one monomer species, so $\kpluseff=\kplusT$, and
$\Dcap(\ccrit) = (\kplusT\ccrit + \vminusT)/2$
\cite{fujiwara:actin_microscopy_large_fluc,ben:living_ionic_letter,ben:living_ionic}.
Using the values of table 1 in these expressions and in eq.
\eqref{cap_crit} gives $\Ncapcrit \approx 26$, of the same order as
the numerics of fig. \ref{cap}.

Finally, note that the preaveraging method shown in fig. \ref{cap} is
an excellent approximation in regions where fluctuations are
unimportant (very large or very small $c$), producing almost identical
results to MC. However, below $\ccrit$ it considerably underestimates
cap lengths.  This results from the preaveraged treatment of
fluctuations.

\section*{\large  Mean Growth Rate, $j(c)$}

\begin{figure}[tb]
\centering
\includegraphics[width=7cm]{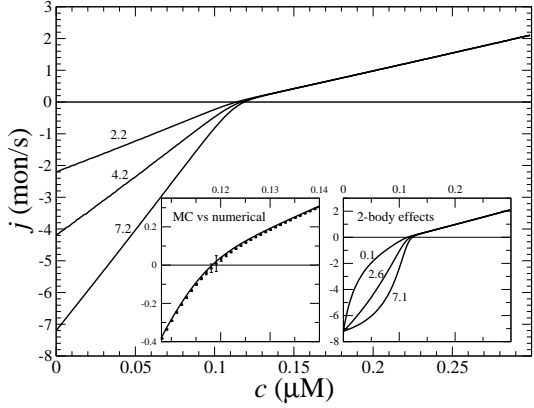}
\caption{\label{jc_ADP} \footnotesize 
Dependence of growth rate on concentration; Influence of $\vminusD$
(indicated in s$^{-1}$ next to each curve).  Other parameters as in
table 1.  MC and exact numerical solution results are
indistinguishable.  The spread in $\ccrit$ values for the 3 curves is
5\%.  Left inset: blow up of critical region showing the agreement
between MC (squares, error bars are standard deviation of mean) and
numerical method (solid line).  Right inset: Influence of many body
effects; the value shown in s$^{-1}$ next to curves is
depolymerization rate of ATP-actin next to ADP-actin, $\vminusTD$.}
\end{figure}

How is the behavior of the average rate of growth $j(c)$ correlated to
cap structure and dynamics?  The lowest curve of fig. \ref{jc_ADP}
shows numerical results for barbed end growth, using identical
parameters to those of fig. \ref{cap}.  A noticeable feature is that
the slopes are very different above and below the critical
concentration of the barbed end.  This directly reflects the cap
structure just discussed, as follows.  For $c\gg \ccrit$ the ATP-actin
segment is long and hides the remaining ADP-Pi-actin portion of the
cap, so $j \approx \kplusT c - \vminusT$ has simple linear form and
slope $\kplusT$, approximately behaving as if ATP-actin were the only
species involved.  On the other hand for $c < \ccrit$, the slope of
$j(c)$ is large because the cap length is changing rapidly as
concentration increases (see fig.  \ref{cap}).  Filament length change
is now generated by capless episodes, when the ADP-actin core is
exposed and the filament shrinks with velocity $\vminusD$ (the steady
state cap has fixed mean length and does not on average
contribute). Thus $j = - \vminusD \pcore$ where $\pcore\approx
1/\Ncap$ is the probability the cap length vanishes, assuming a broad
distribution of cap lengths with mean $\Ncap$.  Using eq.
\eqref{simple}, this gives $j \approx \vminusD \kpluseff (\ccrit -
c)/(2 \Dcap)$ in the region where eq.  \eqref{simple} is valid.  Since
$\vminusD$ is large, this is a much larger slope than for
concentrations above $\ccrit$.

\begin{figure}[tb]
\centering
\includegraphics[width=7cm]{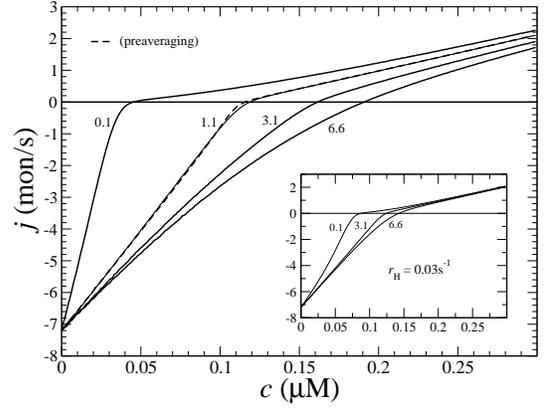}
\caption{\label{jc_ADPPi} \footnotesize 
Growth rate: influence of the value of $\vminusP$ (shown in s$^{-1}$).
Other values as in table 1.  Solid lines: numerical solutions and MC
simulations (indistinguishable).  Dashed line: preaveraging
approximation for $\vminusP=1.1$s$^{-1}$. Inset: same but with
$\rH=0.03$s$^{-1}$.  }
\end{figure}

The region very close to $\ccrit$, where eq. \eqref{cap_crit} takes
over, is an interesting one: (1) Here the total cap becomes long, of
length approximately $\Ncapcrit$, implying that ADP-actin is rarely
exposed at the tip.  This in turn implies that the depolymerization
rate of ADP-actin will have only a small influence on the value of
$\ccrit$.  This is verified in fig.  \ref{jc_ADP} where we display
$j(c)$ curves for $\vminusD$ values ranging from 2.2 to 7.2s$^{-1}$.
These changes produce only a very small shift in $\ccrit$, even though
$j(c)$ changes significantly for $c<\ccrit$.  (2) The mean ATP-cap
length is small (of order unity), and since the tip composition and
cap length is constantly fluctuating, both ATP-actin and ADP-Pi-actin
are frequently exposed at the tip.  Thus we expect a dependence of
$\ccrit$ on the value of $\vminusP$.  This is verified in fig.
\ref{jc_ADPPi} where we display how the growth rate and $\ccrit$
change with the value of $\vminusP$.  The magnitude of the shift is
influenced by the assumed rate of ATP hydrolysis: if one uses, for
example, a hydrolysis rate 10 times smaller, the change in growth
remains substantial but is considerably reduced (see inset).

Note also that preaveraging estimates the growth rate very accurately
(see fig. \ref{jc_ADPPi}). Even in the fluctuation-dominated region
just below $\ccrit$, where cap size is substantially underestimated,
it remains accurate though slightly less so than elsewhere.

\begin{figure}[tb]
\includegraphics[width=6cm]{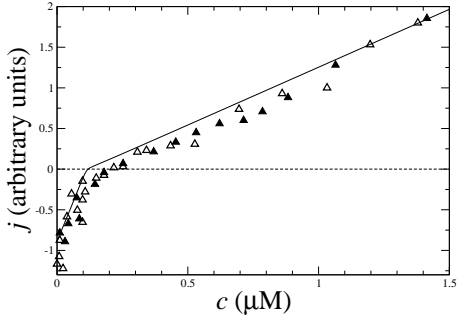}
\centering
\caption{\label{jc_exp} \footnotesize Growth rate $j(c)$ versus
concentration from data taken from fig. 1 of ref.
\citenum{carlier:effect_of_Mg} for simultaneous growth at both ends
(in KCl and Mg).  Solid line: numerical results, barbed end
(parameters from table 1), multiplied by a prefactor to fit data which
lack absolute scale.  Differences between numerical and experimental
results may originate from the pointed end contribution, or possibly
due to the experimental ionic conditions.
}
\end{figure}

An important question is the effect of many body interactions between
actin subunits, so far neglected in this report.  We have found that
the shape of the mean growth rate near and below the critical
concentration is sensitive to these.  As an example, fig. \ref{jc_ADP}
(inset) shows the dependence of $j(c)$ on the depolymerization rate of
ATP-actin when its nearest neighbor is ADP-actin ($\vminusTD$), with
all other rates as in table 1.  Other types of many body interactions
can lead also to shifts in $\ccrit$ (not shown).  Including many body
interactions rapidly increases the number of rate constants.  Since
these are unknown and presumably hard to measure, this limits the
uniqueness with which growth rate curves can be modeled near
$\ccrit$.  We stress, however, that the central qualitative
conclusions, namely the existence of a long cap at $\ccrit$ and the
associated change of slope of the growth rate, are general.  An
example of fitting experimental $j(c)$ curves with a one body model is
shown in fig. \ref{jc_exp}.


\section*{\large Fluctuations in Growth Rate}

\begin{figure}[tb]
\centering
\includegraphics[width=7.5cm]{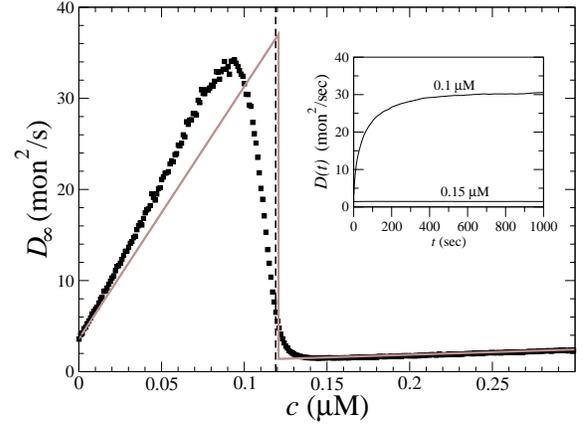}
\caption{\label{dc} \footnotesize 
Long time length diffusion coefficient, $\Dinf(c)$.  Squares: MC
results, parameters from table 1.  Vertical dashed line indicates
$\ccrit$.  Solid line: prediction of simple model:
$\vminusP=\vminusT=1.4$s$^{-1}$, $\rPi=0$, other values from table 1.
Inset: Time-dependence of diffusivity at two concentrations.}
\end{figure}

Turning now to fluctuations in growth rates, we find these behave
dramatically around the critical concentration reflecting a mild
version of the dynamic instability exhibited by microtubules
\cite{hill:aggr_bio_book,hillchen:microtubule_phase_change_84}.  In
the inset of fig. \ref{dc} we used MC to evaluate the length
diffusivity, $D(t) \equiv (\av{L^2}-\av{L}^2)/(2t)$, where $L$ is the
number of subunits added/subtracted after time $t$, starting from
filaments with steady state caps at $t=0$.  For $c=0.15\mu$M (above
$\ccrit$) we find $D$ is essentially independent of time.  Its
magnitude is of order 1mon$^2$/s, as would be expected for a growth
process of identical subunits which add/subtract with rates of order
1s$^{-1}$
\cite{fujiwara:actin_microscopy_large_fluc,hill:aggr_bio_book,%
ben:living_ionic_letter,ben:living_ionic}.  However, for $c = 0.1\mu$M
(below $\ccrit$) $D$ is {\em increasing} with time, reaching a large
asymptotic value $\Dinf$ after several hundred seconds.  Fig. \ref{dc}
shows the dependence of $\Dinf$ on concentration; it exhibits a sharp
peak below $\ccrit$ and then drops rapidly.

To understand the physics underlying this behavior, consider the
simple model where ATP-actin and ADP-Pi-actin are identical
($\vminusT=\vminusP$) and Pi release very slow ($\rPi \gt 0$).  Now
$D$ describes the random walk performed by the filament tip; if the
tip makes a random forwards or backwards step of $L$ monomer units
every time interval $T$, then one can write $D =L^2/T$.  Just above
the critical concentration, where on and off rates are approximately
equal, the tip randomly adds or subtracts one ATP-actin ($L=1$) in a
mean time $T=1/\vminusT$, giving $D = \vminusT$.  Just below the
critical concentration, however, we know there is a long steady state
cap.  Since most filaments are capped, at short times $D$ is
determined by length changes of the cap and its value is thus close to
the cap diffusivity, $\Dcap$.  As time increases, more and more
uncapping episodes occur, each episode now contributing to filament
length change.  Such events are correlated on the timescale of the cap
lifetime, $\tcap \approx \Ncap^2/\vminusT$ (we used $\Dcap=\vminusT$
for the simple model \cite{ben:living_ionic_letter,ben:living_ionic}.)
This explains why $D(t)$ changes with time up to the cap lifetime (see
fig. \ref{dc}, inset).  Thus to determine $\Dinf$, one must take
$T=\tcap$.  Using a well known result from the theory of 1D random
walks\cite{fisher:walks_jstatphys_84}, the number of uncapping events
during the time $\tcap$ is approximately $(\Dcap \tcap)^{1/2} \approx
\Ncap$.  Since the number of core monomers lost during each uncapping
episode before a polymerizing monomer arrives is of order
$\vminusD/\vminusT$, thus $L = \Ncap \vminusD/\vminusT$.  This leads
to a very different expression for the diffusivity, $\Dinf \approx
(\vminusD)^2/\vminusT$: there is a discontinuity in diffusivity at
$\ccrit$ of magnitude
                                                \begin{eq}{door}
\Delta \Dinf = \vminusT (\lambda^2 - 1) 
\comma  \ \ \ \
\lambda \equiv \vminusD/\vminusT \period
\end{eq}
At the barbed end the instability parameter $\lambda \approx 5.1$ and
fluctuations at the critical concentration are very large, with a
pronounced discontinuous drop in $\Dinf$ as one passes to higher $c$.
A rigorous derivation of eq. \eqref{door} is shown in the
supporting material where in addition we obtain the full sawtooth
curve shown in fig. \ref{dc}; evidently, the simple model captures
many features of the actual $\Dinf(c)$ profile. The effect of Pi
release and ATP-actin/ADP-Pi-actin differences is to shift $\ccrit$
and to smooth the sharp peak and shift it to somewhat below $\ccrit$.

How do the results of fig.  \ref{dc} compare to the large fluctuations
observed by Fujiwara et al.
\cite{fujiwara:actin_microscopy_large_fluc} and Kuhn and Pollard
\cite{kuhnpollard:tirf_bj_04}, and also suggested by the findings of
ref.  \citenum{brennerkorn:actin_subunit_exchange}?  Fig.  \ref{dc}
shows a peak value of $\Dinf\approx 34$mon$^2$s$^{-1}$, dropping to
$\Dinf\approx 5$mon$^2$s$^{-1}$ at $\ccrit$.  The experimentally
reported value was $\twid 30$mon$^2$s$^{-1}$; however, these
measurements were performed at
\cite{fujiwara:actin_microscopy_large_fluc} or close
\cite{kuhnpollard:tirf_bj_04} to a treadmilling steady state, \ie at a
concentration slightly above $\ccrit$ for the barbed end and well
below that for the pointed end.  At this concentration, fig.  \ref{dc}
shows a diffusivity less than $5$mon$^2$s$^{-1}$.  Thus both theory
and experiment exhibit large fluctuations near $\ccrit$, but at
different concentrations.  Future experimental measurements of the
full $\Dinf(c)$ profile are needed to establish the relationship, if
any, between these.

Our work leads also to the following prediction: since phosphate will
bind to ADP-actin and eliminate the effect of a large instability
parameter, thus fluctuations and $D$ at the barbed end will be
suppressed in the presence of excess Pi.


\section*{\large  Discussion}

{\bf Pointed End $j(c)$: Why is $\ccrit$ so Different?}  In this study
we emphasized the barbed end, but our methods are also applicable to
the pointed end, provided the same mechanisms of uniform random
hydrolysis and slow Pi release remain valid.  Making this assumption,
let us now discuss why $\ccrit$ (for ATP-actin) at the pointed end is
almost six times the value at the barbed end
\cite{pollard:actin_rate_constants}. Now an important issue is how
different the ATP-actin and ADP-Pi-actin species are, in terms of on
and off rate constants.  That they are similar is suggested by the
observation that excess phosphate reduces the critical concentration
in a pure ADP-actin polymerization to a value rather close to the
barbed end $\ccrit$ in ATP
\cite{rickardsheterline:pi_affects_adp_critical_conc,
wangerwegner:pi_binding_actin,carlier:adppi_role,weber:actin_tropomodulin_adppi_ccrit_jbc_99}.
But if indeed the 2 species are similar, and the same basic mechanisms
apply at the pointed end, this is inconsistent with the very different
$\ccrit$ values. This inconsistency is due to the cap structure we
have established here: the cap includes a long ADP-Pi segment
essentially {\em hiding} the ADP-actin core which is thus {\em rarely
seen} at the filament tip (see fig.  1(a)).  For the barbed end (fig.
2) $\Ncap \approx 25$ at $\ccrit$, and we find a large value for the
pointed end at its $\ccrit$, though smaller than the barbed end (data
not shown).  Thus ADP-actin on/off rates are almost irrelevant to
$\ccrit$ (see fig. 3) and hence differences between ATP-actin and
ADP-actin cannot account for the large $\ccrit$ differences.  Thus the
origin must be different ATP-actin/ADP-Pi-actin compositions at the
pointed and barbed ends; since the ATP-actin segment is short both
species are regularly exposed at filament ends and substantially
different $\ccrit$ values will result provided the 2 species have
different rate constants.  Were these identical, $\ccrit$ at both ends
would be very similar, because the on/off rates at the filament ends
would then be very close to the values for an all ATP-actin filament;
for such a filament, detailed balance dictates that the ratio of
on/off rates at each end are identical \cite{hill:aggr_bio_book}.
(However, in apparent contradiction to this conclusion are the
findings of ref.  \cite{pollard:actin_rate_constants} where different
on/off ratios were reported at each end, under conditions where long
ATP-actin caps are expected. A conceivable explanation is possibility
(ii), see below.)  Many body effects will further affect $\ccrit$.

We are driven to two possibilities: (i) ATP-actin and ADP-Pi-actin are
substantially different, or (ii) different mechanisms operate during
pointed end growth.  Certain workers
\cite{carlier:atp_hydro_irreversible,dickinson:force_endtrack_bj_04}
have proposed possibility (i), based on the irreversibility of
hydrolysis \cite{carlier:atp_hydro_irreversible} which suggests a
large energetic change, possibly a structural change of the filament.
Possibility (i) may in fact be consistent with the experiments of
refs.  \citenum{rickardsheterline:pi_affects_adp_critical_conc,
wangerwegner:pi_binding_actin,carlier:adppi_role,weber:actin_tropomodulin_adppi_ccrit_jbc_99}
which did not probe individual on/off rate constants of ADP-Pi-actin
and which may have involved significant ADP-actin polymerization
\cite{wangerwegner:pi_binding_actin}.  We are unaware of any
crystallographic \cite{otterbein:uncomplexed_actin_structure} or
electron microscopic \cite{belmont:actin_conformation_pi_release}
experiments examining ATP/ADP-Pi differences for filamentous actin.

If we adhere to the assumption that the growth mechanisms as
previously outlined apply to both ends, we are then led to the
following prediction: the values of $\ccrit$ for ATP-actin at both
ends will be only weakly affected by the presence of excess Pi
(provided ionic conditions are strictly unchanged).  This is because
the binding of Pi to ADP-actin segments is almost irrelevant since
these are rarely exposed at the tip due to long caps at $\ccrit$.
Indeed, for the barbed end no significant shift has been observed in
the presence of Pi
\cite{rickardsheterline:pi_affects_adp_critical_conc,
wangerwegner:pi_binding_actin,carlier:adppi_role,weber:actin_tropomodulin_adppi_ccrit_jbc_99}.
For the pointed end, however, a reduction of $\ccrit$ has been
reported in the presence of Pi and barbed end capping proteins
\cite{rickardsheterline:pi_affects_adp_critical_conc,
wangerwegner:pi_binding_actin,carlier:adppi_role,weber:actin_tropomodulin_adppi_ccrit_jbc_99}.
This cannot be explained within the present framework and suggests
possibility (ii).  Future experiment will hopefully settle this
important issue.

{\bf Conclusions.}  In this work filament growth rates $j(c)$ and
their fluctuations, as measured by the diffusivity $D(c)$, were
calculated as functions of ATP-actin concentration $c$.  To our
knowledge, this is the first rigorous calculation of these quantities
accounting for all known basic mechanisms.  Pantaloni et al.
\cite{pantaloni:hill_actin_atp_cap_1,hill:actin_atp_cap_2} studied
$j(c)$ at the barbed end in a work before the mechanism of Pi release
was discovered.  Infinitely fast Pi release and vectorial hydrolysis
were assumed.  Given the data available at that time, in order to
explain the sharp change in slope of $j(c)$ at $\ccrit$ (see \eg
fig. 5), they further assumed (i) strong three body
ATP-actin/ADP-actin interactions which lead to stable short ATP-actin
caps, and (ii) zero hydrolysis rate of the nucleotide bound to the
terminal monomer.  In our work, the origin of the sharp change in
slope is precisely the fact that Pi release is slow, similarly to an
earlier model of microtubule polymerization
\cite{hillcarlier:gtp_hydro_steady_state}.

Recently, Bindschadler et al.
\cite{bindschadler:actin_mechanistic_model} studied the composition of
actin filaments accounting for all three actin species at steady
state.  We have examined the preaveraging approximation used in their
work and showed that it leads to very accurate $j(c)$ curves, but the
cap lengths are underestimated below $\ccrit$.

Here we have addressed random ATP hydrolysis only.  Future work is
needed to analyze the implications of the vectorial hydrolysis
suggested by refs.  \citenum{carlier:hydrolysis_mechanism,
pantaloni:hill_actin_atp_cap_1}.  We showed that for random hydrolysis
$j(c)$ is linear far above the critical concentration.  Growth rate
experiments for both ends together in the absence of KCl have
exhibited non-linearities up to $c=10\mu$M, far above the critical
concentration of the barbed end which is 1$\mu$M under these
conditions \cite{carlier:atp_cap_evidence,carlier:adp_atp_sonication}.
In refs.
\citenum{carlier:atp_cap_evidence,pantaloni:hill_actin_atp_cap_1} this
observation was attributed to vectorial hydrolysis at the barbed end
while in ref.  \citenum{pollard:actin_rate_constants} this was
assigned to the non-linear contribution of the pointed end whose
critical concentration is $\approx 5\mu$M under the same conditions.

Perhaps our most interesting finding is that the long time diffusivity
$\Dinf$ has a large peak below the critical concentration $\ccrit$ of
the barbed end, followed by a sharp drop in a narrow range above
$\ccrit$.  This conclusion is quite general and its origin is the
smallness of the Pi release rate and the large value of the off rate
of ADP-actin at the barbed end.
Future measurements of length diffusivities over a range of
concentrations promise to provide new information and insight on the
fundamentals of actin polymerization.

{\small

This work was supported by the Petroleum Research Fund, grant
33944-AC7, and NSF, grant CHE-00-91460.  We thank Ikuko
Fujiwara, Jeffrey Kuhn and Thomas Pollard for stimulating
discussions.}



{\footnotesize

}

\pagebreak


\onecolumn
\section*{SUPPORTING MATERIAL}

\newcommand{\von}{\kplusT c}
\newcommand{\bzero}{b_0}
\newcommand{\bone}{b_1}
\newcommand{\btwo}{b_2}
\newcommand{\Estar}{E^*}



\subsection*{A. Numerical Method for Growth Rates.}

In this part of the supporting material we describe the numerical
method we used in the main text to calculate the growth rate curves of
figs.  \ref{jc_ADP} and \ref{jc_ADPPi}.  Consider an ATP-actin monomer
which polymerizes at the filament tip at time $t=0$.  We define the
return probabilities $\psiT_t$, $\psiP_t$, and $\psiD_t$ to be the
probability that this monomer is once again at the tip after time $t$
as ATP-actin, ADP-Pi-actin, or ADP-actin, respectively.  The total
depolymerization rate of this monomer at time $t$ is
                                                \begin{eq}{totald}
F_t  =  \vminusT\psiT_t + \vminusP\psiP_t + \vminusD\psiD_t \period
\end{eq}
The dynamical equations obeyed by the return probabilities are
            \begin{eqarray}{evolution}
\ddt\psiT_t & = & -(\von + \vminusT +\rH)\psiT_t
        + \von \int_0^t dt' \psiT_{t'} F_{t-t'} e^{-\rH(t-t')}
                                               \comma\drop
\ddt\psiP_t & = & -(\von + \vminusP + \rPi)\psiP_t
        + \rH\psiT_t
        + \von \int_0^t dt' \psiP_{t'} F_{t-t'} e^{-\rPi(t-t')}
                                               \drop
        &+& \von \int_0^t dt' \psiT_{t'} F_{t-t'} 
                {\rH \over \rPi-\rH}
                \paren{e^{-\rH(t-t')} - e^{-\rPi(t-t')}}
                                               \comma\drop
\ddt\psiD_t & = & -(\von + \vminusD)\psiD_t
        + \rPi\psiP_t
        + \von \int_0^t dt' \psiD_{t'} F_{t-t'}
        + \von \int_0^t dt' \psiP_{t'} F_{t-t'} 
                \paren{1 - e^{-\rPi(t-t')}}
                                               \drop
        &+& \von \int_0^t dt' \psiT_{t'} F_{t-t'} 
          \paren{1 - {\rPi \over \rPi-\rH}e^{-\rH(t-t')}
                 +{\rH \over \rPi-\rH}e^{-\rPi(t-t')}}
                                                \period
\end{eqarray}
Here the non-integral terms on the right hand sides represent change
of tip status due to polymerization, depolymerization, hydrolysis, and
phosphate release events at time $t$.  The integral terms represent
rates of reappearance of the monomer at the tip, weighted by factors
accounting for the probability of hydrolysis or phosphate release
during the time interval since the last appearance at the tip.  For
example, the first term on the right hand side of the first equation
represents the rate of change of the probability of finding the
ATP-actin monomer at the tip due to (i) polymerization of another
monomer on top of it, (ii) depolymerization of the monomer itself, or
(iii) hydrolysis of the ATP nucleotide bound to the monomer at the
tip.  The integral term on the right hand side represents reappearance
events of the ATP-actin unit at the tip given that it got buried
inside the filament due to a polymerization event at time $t'$, an
event which occurred with rate $\von$.  Factor $F$ represents the rate
of reappearance of the buried monomer at the tip due to
depolymerization of all the monomers which were added on top of it.
The factor $e^{-\rH(t-t')}$ is the probability that the ATP-actin
monomer in question is not hydrolyzed while being buried.

Now the filament growth rate is given by
                                                \begin{eq}{fox}
j = \casesbracketsii
{\vminusD \square{1-\int_0^\infty dt(\psiD_t + \psiP_t +\psiT_t)}}{j < 0} 
{\kplusT c -\int_0^\infty dt F_t} {j > 0}
\end{eq}
Carrying out a Laplace transformation, $t \gt E$, $F_t \gt
f_E$, and $\psi_t \gt \Psi_E$ one has from eq. \eqref{fox}
                                                \begin{eq}{dog}
j = \casesbracketsii
{\vminusD \square{1- \PsiD_0 - \PsiP_0 -\PsiT_0}}{j < 0} 
{\kplusT c -f_0} {j > 0}
\end{eq}
while from eq. \eqref{evolution} one obtains
                                                \begin{eqarray}{revolution}
\PsiT_E &=& 1/\paren{E+\vminusT+\rH+\von(1-f_{E+\rH})}
                                                \comma\drop
\PsiP_E &=& {\rH + \von\ \rH (f_{E+\rH}-f_{E+\rPi}) / (\rPi-\rH)
        \over E+\vminusP+\rPi+\von(1-f_{E+\rPi})} \PsiT_E
                                                \comma\drop
\PsiD_E &=& {\paren{\rPi + \von(f_E - f_{E+\rPi})} \PsiP_E
                +\von\paren{f_E - \rPi f_{E+\rH}/(\rPi-\rH) +\rH
                        f_{E+\rPi}/(\rPi-\rH) }\PsiT_E 
                \over
             E + \vminusD + \von(1-f_E)}
                                                \period
\end{eqarray}
Eliminating all $\Psi$ in the Laplace transformation of
eq. \eqref{totald} after using eq. \eqref{revolution} one obtains the
following recursive relationship involving the function $f$ alone:
                                                \begin{eq}{salmon}
f_E = {\cal R}[f_{E+\rH},f_{E+\rPi}]\comma \gap
\end{eq}
where
                                                \begin{eq}{recursive}
{\cal R}[f_{E+\rH},f_{E+\rPi}] = {-\bone + \sqrt{\bone^2 - 4\btwo\bzero}\over 2\btwo}
                        \period
                                                                \end{eq}
Here the symbols $\bzero$, $\bone$, and $\btwo$ are functions of
$E$, $f_{E+\rH}$ and  $f_{E+\rPi}$ as follows:
                                                \begin{eqarray}{blue}
\bzero &=& A_{0,2} E^2 + A_{0,1} E + A_{0,0}
                                        \comma\drop
\bone &=& A_{1,3} E^3 + A_{1,2} E^2 + A_{1,1} E + A_{1,0} 
                                        \comma\drop
\btwo  &=&  A_{2,2} E^2  +  A_{2,1} E  +  A_{2,0}
\comma
                                                                \end{eqarray}
where
                                                \begin{eqarray}{bubble}
A_{0,2} &=& -(\rH - \rPi) \vminusT \von
                                        \comma\drop
A_{0,1} &=& (\vminusD \rH - \rPi \vminusT + \rH(-\vminusP + \vminusT))(\von)^2 f_{E+\rPi}
        +(\rH \vminusP -\vminusD \rPi)(\von)^2 f_{E+\rH}
                                        \drop
        &-&(\rH - \rPi) \von (\rH \vminusP + \vminusT(\vminusD+\vminusP+\rPi+2\von))
                                        \comma\drop
A_{0,0} &=& - \vminusD (\rH-\rPi)(\von)^3 f_{E+\rPi} f_{E+\rH}
                                        \drop
        &-&(\rH(\vminusP-\vminusT)+\rPi \vminusT)\von+\vminusD((\rH)^2-\rPi \vminusT +\rH(-\rPi +\vminusT+\von))(\von)^2 f_{E+\rPi}
                                        \drop
        &+&(\rH \vminusP \von +\vminusD(\rH(\vminusP+\rPi)-\rPi(\vminusP+\rPi+\von))) (\von)^2 f_{E+\rH}
                                        \drop
        &-&(\rH-\rPi)\von(\von(\rH \vminusP+\vminusT(\vminusP+\rPi+\von))+\vminusD(\rH(\vminusP+\rPi)+\vminusT(\vminusP+\rPi+\von)))
                                        \comma\drop
A_{1,3} &=& \rH - \rPi
                                        \comma\drop
A_{1,2} &=& -(\rH-\rPi)\von (f_{E+\rPi}+f_{E+\rH})
        +(\rH-\rPi)(\vminusD+\rH+\vminusP+\rPi+\vminusT+3\von)        
                                        \comma\drop
A_{1,1} &=& (\rH-\rPi)(\von)^2 f_{E+\rPi} f_{E+\rH}
        -(\rH-\rPi)(\vminusD+\rH+\vminusT+2\von)\von f_{E+\rPi}
                                        \drop
        &-&(\rH-\rPi)(\vminusD+\vminusP+\rPi+2\von)\von f_{E+\rH}
                                        \drop
        &+&(\rH-\rPi)(\vminusP \vminusT+\rPi \vminusT+2\vminusP \von+2\rPi
\von+3 \vminusT \von + 3(\von)^2
                                        \drop
        &&\gap +\vminusD(\rH+\vminusP+\rPi+\vminusT+\von)+\rH(\vminusP+\rPi+2\von))
                                        \comma\drop
A_{1,0} &=& (\rH-\rPi)(\vminusD+\von)(\von)^2 f_{E+\rPi} f_{E+\rH}
                                        \drop
        &+&(-\vminusD(\rH^2 - \rPi \vminusT + \rH(-\rPi+\vminusT+\von)) 
                                        \drop
        &&\gap  +\von(-\rH^2+\rH(\vminusP+\rPi-2\vminusT-\von)+\rPi(2\vminusT+\von))) \von f_{E+\rPi}
                                        \drop
        &+&(\vminusD(-\rH(\vminusP+\rPi)+\rPi(\vminusP+\rPi+\von)) 
                                        \drop
        &&\gap  +\von(\rPi(\vminusP+\rPi+\von)-\rH(2\vminusP+\rPi+\von))) \von f_{E+\rH}
                                        \drop
        &+&(\rH-\rPi)(\vminusD(\rH(\vminusP+\rPi)+\vminusT(\vminusP+\rPi+\von))
                                        \drop
        &&\gap + \von(\rH(2\vminusP+\rPi+\von)+(\vminusP+\rPi+\von)(2\vminusT+\von)))
                                        \comma\drop
A_{2,2} &=& \rPi - \rH
                                        \comma\drop
A_{2,1} &=& (\rH - \rPi)\von(f_{E+\rPi}+f_{E+\rH})
        -(\rH-\rPi)(\rH+\vminusP+\rPi+\vminusT+2\von)
                                        \comma\drop
A_{2,0} &=& -(\rH-\rPi)(\von)^2 f_{E+\rPi} f_{E+\rH}
        +(\rH-\rPi)(\rH+\vminusT+\von)\von f_{E+\rPi}
                                        \drop
        &+&(\rH-\rPi)(\vminusP+\rPi+\von)\von f_{E+\rH}
        -(\rH-\rPi)(\vminusP+\rPi+\von)(\rH+\vminusT+\von)
                                        \period
\end{eqarray}
We remark that eq. \eqref{recursive} is the solution of a quadratic
equation; which of the two solutions of the quadratic is meaningful is
easily checked by demanding $f < 1$ in the $E\gt\infty$ limit.

Now for any given monomer concentration $c$, with the boundary
condition $f_E \gt 0$ as $E\gt\infty$, we started from a large enough
$E$ value and evolved eq.  \eqref{salmon} towards $E =0$ to obtain
$f_0$, $f_{\rPi}$, and $f_{\rH}$.  Substituting these values in eq.
\eqref{revolution} we further obtained $\PsiT_0, \PsiP_0$ and
$\PsiD_0$.  Thus, given $f_0, \PsiT_0, \PsiP_0$ and $\PsiD_0$ we
evaluated $j(c)$ using eq. \eqref{dog}.  It was shown that this method
converges to a unique solution provided one starts the evolution from
large enough $E$, retaining a sufficient number of significant digits.

\ignore{
The solution of the recursive relationship involves the following
technical points:

(i) Must choose large enough $E$   (*)

(ii) Must use many significant digits (convergence problem at small $E$).

(iii) What is the initial condition exactly (*)
} 

\subsection*{B. Preaveraging Approximation: Calculation of Growth Rates.}

In this part of the supporting material we present the method we used
to calculate cap sizes and growth rates based on a preaveraging
approximation.  As discussed in the main text, compared to the exact
calculations, this method gives different results for the cap size
below $\ccrit$, but provides very good approximations to growth rates.
We denote $\phiT(n)$, $\phiP(n)$ and $\phiD(n)$ the probability that
the $n^{\rm th}$ monomer away from the tip binds ATP, ADP-Pi or ADP,
respectively.  Consider first the tip, \ie $n=1$.  One
has\cite{bindschadler:actin_mechanistic_model}
                                                \begin{eq}{t_one}
\ddt \phiT(1) 
= 
\kplusT c [\phiP(1)+\phiD(1)]
+
[\vminusP \phiP(1)+\vminusD \phiD(1)]\phiT(2)
-
\vminusT [\phiP(2)+\phiD(2)] \phiT(1)
-
\rH \phiT(1) \period
\end{eq}
The first term on the right hand side represents change of tip status
into ATP-actin due to polymerization of ATP-actin at an ADP-Pi-actin
or ADP-actin tip.  The second term represents change of tip status
into ATP-actin due to (i) depolymerization of ADP-Pi-actin or
ADP-actin at $n=1$, and (ii) exposure of ATP-actin, previously buried
at position $n=2$.  Within the preaveraging approximation, the joint
probability of ADP at $n=1$ and simultaneously ATP at $n=2$, for
example, is approximated as a product of probabilities: $\phi_{\rm
DT}(1,2) \approx \phiD(1)\, \phiT(2)$ in eq. \eqref{t_one}.  Similarly,
the third term on the right hand side represents depolymerization of
ATP-actin while the last term is change due to hydrolysis.

For ADP-Pi-actin one has similarly
                                                \begin{eq}{p_one}
\ddt \phiP(1) 
= 
[\vminusT \phiT(1)+\vminusD \phiD(1)]\phiP(2)
-
\curly{\vminusP[\phiT(2)+\phiD(2)]+\kplusT c + \rPi} \phiP(1)
+ 
\rH \phiT(1) \period
\end{eq}
The analogous equations for $n>1$ are
\cite{bindschadler:actin_mechanistic_model} 
                                                \begin{eqarray}{ptn}
\ddt \phiT(n) 
&=& 
\kplusT c [\phiT(n-1)-\phiT(n)]
+
[\vminusT \phiT(1) + \vminusD \phiD(1) +\vminusP \phiP(1)][\phiT(n+1)-\phiT(n)]
-
\rH \phiT(n)
\ddrop
\ddt \phiP(n) 
&=& 
\kplusT c [\phiP(n-1)-\phiP(n)]
+
[\vminusT \phiT(1) + \vminusD \phiD(1) +\vminusP \phiP(1)][\phiP(n+1)-\phiP(n)]
\ddrop
&+&
\rH \phiT(n)
-
\rPi \phiP(n) 
\ggap \ggap  (n>1)
\end{eqarray}
The rate of change of the ADP-actin probabilities are determined from
$\phiT(n)+\phiP(n)+\phiD(n) = 1$.  Starting from an arbitrary
nucleotide profile and using long filaments, we evolved numerically
eqs.  \eqref{t_one}-\eqref{ptn} for a long enough time to allow the
profile to reach its steady state (note that an analytical solution is
also possible since \eqref{ptn} is linear, except for the tip terms
\cite{bindschadler:actin_mechanistic_model}).  The growth rate and the
cap size were calculated from
                                                \begin{eq}{element}
\Ncap = \sum_{n=1}^{\infty} \phiT(n)+\phiP(n) \comma \gap
\NATPcap = \sum_{n=1}^{\infty} \phiT(n) \comma \gap
j = \kplusT c - \vminusT \phiT(1) - \vminusP \phiP(1) - \vminusD \phiD(1)
\period
\end{eq}

\subsection*{C. Analytical Calculation of Length Diffusivity as a
Function of Concentration.}

Here we prove that $\Dinf(c)$ in fig. \ref{dc} has a sawtooth shape
in the special case $\vminusP=\vminusT$ and $\rPi\gt 0$.  Consider
first shrinking barbed ends, $c<\ccrit$.  For long times, $t \gg
\tcap$, apart from fluctuations in the size of the steady state
ATP/ADP-Pi cap, fluctuations in tip displacement are due to
fluctuations in how far the ADP core has shrunk.  Let us call
$u(\tau|t)$ the probability that in time $t$ the filament is uncapped
for a total time $\tau$.  The probability that $N$ monomers have been
lost during this time is
                                                \begin{eq}{random}
p(N,t) = \int_0^\infty d\tau\, u(\tau|t)\,
\Poisson(N,\vminusD \tau) \comma \gap
\Poisson(N,x) \equiv x^N e^{-x}/N! \comma
\end{eq}
where the Poisson distribution, $\Poisson$, describes the probability
distribution of the number of depolymerized ADP-actin monomers during
the total uncapped period.  Let us evaluate $u$ by noting that the
average value of $\tau$ and its second moment are
given by
                                                \begin{eq}{second}
\av{\tau} = t\ \Sinf \comma \gap
\av{\tau^2} = 2 \Sinf \int_0^t dt'
\int_{t'}^t S(t'' - t') dt'' \comma \
\end{eq}
for long enough times. Here $S(t)$ is a return probability, namely the
probability that the tip is uncapped at time $t$, given that it was
uncapped at $t=0$, and $S(t)\gt \Sinf$ for $t\gt \infty$.  To prove
eq. \eqref{second}, define $\xi(t''|t')$ to be a random variable which
is zero (unity) when the tip is capped (uncapped) at time $t''$, given
it was uncapped at $t'$.  One has $\av{\tau^2} = \int_0^t\int_0^t dt'
dt'' \av{\xi(t'|0) \xi(t''|0)} = 2\int_0^t dt' \int_{t'}^t dt''
\av{\xi(t'|0)} \av{\xi(t''|t')}$.  Noting that $S(t) = \av{\xi(t|0)}$
one recovers eq. \eqref{second}.

Now the exact result for $S$ is \cite{hill:aggr_bio_book} 
                                                \begin{eq}{complicated}
S(t) = e^{-(\kplusT c + \vminusT) t} [ I_0 (2 t x) + y^{-1/2}
I_1(2 t x) + (1-y) \sum_{j=2}^\infty y^{-j/2} I_j(2 t x)]\comma 
\ \ \ 
y \equiv \kplusT c /\vminusT \comma \ \ \
 x \equiv (\kplusT c
\vminusT)^{1/2} \comma
\end{eq}
where $I_j$ are modified Bessel functions.  Using $\Sinf = 1-c/\ccrit$
one obtains for long times $\av{\tau} = t (1-c/\ccrit)$ and
$\av{\tau^2}_{\rm c} = (t/ \vminusT)$ where $\av{}_c$ denotes second
cumulant.  Thus relative fluctuations in $\tau$ become small for long
times and $u$ becomes Gaussian, $u(\tau|t) \approx {\rm const.}\ 
\exp[-(\tau - \av{\tau})^2/(2 \av{\tau^2}_c)]$.  Substituting $u$ in
$p(N,t)$ of eq. \eqref{random} and performing the integration one
obtains 
                                                \begin{eq}{whale}
\Dinf(c) = \casesbracketsshortii
{(\vminusD/2) [1 + (2\vminusD/\vminusT - 1) c/\ccrit] }{c<\ccrit}
{(\kplusT c + \vminusT)/2}{c>\ccrit}
\end{eq}
which is the sawtooth curve plotted in fig. \ref{dc}.  Notice that
$\Dinf$ decreases for smaller concentrations since at $c=0$ one must
recover the Poissonian fluctuations of a pure depolymerization process
for which $\Dinf = \vminusD/2$.  In eq. \eqref{whale}, the $c >
\ccrit$ expression represents the fluctuations of a polymerization
process of identical subunits \cite{ben:living_ionic} (since in the
limit considered here the cap is never lost above $\ccrit$).


\end{document}